\documentclass[aps,prl,twocolumn,superscriptaddress,amsfont,graphicx,nofootinbib,preprintnumbers]{revtex4}%
\usepackage{color,graphicx,epsfig}
\usepackage{ifpdf}
\usepackage{amsmath}
\usepackage{bm}
\usepackage{color}
\usepackage[english]{babel}
\usepackage{graphicx}%
\usepackage{amsfonts}%
\usepackage{amssymb}
\usepackage{braket}
\usepackage{hyperref}
\usepackage{enumerate}

\definecolor{nicered}{rgb}{0.7,0.1,0.1}
\definecolor{nicegreen}{rgb}{0.1,0.5,0.1}
\hypersetup{colorlinks,citecolor= nicegreen,linkcolor= nicered}

\newcommand{\be}{\begin{equation}}
\newcommand{\ee}{\end{equation}}
\newcommand{\bea}{\begin{eqnarray}}
\newcommand{\eea}{\end{eqnarray}}

\definecolor{Red}{rgb}{1.,0.,0.}

\def\OMIT#1{}

\begin{document}

\def\JHU{Department of Physics and Astronomy, Johns Hopkins University, Baltimore, USA}
\def\KIT{Institute for Theoretical Particle Physics (TTP), KIT, Karlsruhe, Germany}
\def\CERN{CERN Theory Division, CH-1211, Geneva 23, Switzerland}
\def\IPPP{Institute for Particle Physics Phenomenology, Durham University, 
Durham, DH1 3LE, UK 
}

\preprint{IPPP/17/20, TTP17-012}

\title{Top-bottom interference effects in Higgs plus jet production at the LHC}

\author{Jonas M. Lindert }            
\email[Electronic address:]{jonas.m.lindert@durham.ac.uk}
\affiliation{\IPPP}

\author{Kirill Melnikov}            
\email[Electronic address:]{kirill.melnikov@kit.edu}
\affiliation{\KIT}

\author{Lorenzo Tancredi}            
\email[Electronic address:]{lorenzo.tancredi@kit.edu}
\affiliation{\KIT}

\author{Chris Wever}            
\email[Electronic address:]{christopher.wever@kit.edu}
\affiliation{\KIT}

\begin{abstract}
We compute  next-to-leading order QCD corrections 
to the top-bottom interference  contribution 
to $H+j$ production  at the LHC. 
To achieve this, we combine the recent computation of the two-loop amplitudes for $gg \to Hg$ and $qg \to Hq$,
performed
in the approximation of a small $b$-quark mass, and the numerical 
calculation of the squared one-loop amplitudes for $gg \to Hgg$ and $q g \to H q g$ 
performed within {\tt OpenLoops}.  
We find that QCD corrections to the interference are large 
and similar to the QCD corrections to 
the top-mediated Higgs production cross section.  We also observe a significant reduction 
in the mass-renormalization scheme uncertainty,  once the NLO QCD prediction for the 
interference  is employed. 
\end{abstract}

\maketitle
%%%%%

Detailed exploration of the Higgs boson properties is a major part of the 
physics program at the  Large Hadron Collider (LHC).  It is hoped that studies of the Higgs 
couplings  will reveal  possible physics beyond the Standard Model, especially if it 
mostly manifests itself through interactions with the Higgs bosons.  The goal, therefore, 
is to precisely measure Higgs boson couplings to various particles in  the Standard Model 
and to search for small deviations. For example, assuming that the energy scale of New Physics 
is close to $1~{\rm TeV}$, generic modifications of  the Higgs couplings are
expected at the level of $v^2/(1~{\rm TeV})^2 \sim 5 \times 10^{-2}$, 
where we used $v =  246~{\rm GeV}$ for the Higgs field vacuum expectation value.  
A variety of explicit 
BSM scenarios  conforms with these expectations \cite{Gupta:2013zza}, suggesting that 
achieving a few percent precision in studies of the  Higgs couplings may indeed provide  interesting 
information about physics beyond the Standard Model. 

Compared to these theoretical  goals, existing  measurements of the Higgs couplings 
leave  much to be desired \cite{Khachatryan:2016vau}.
Currently,  Higgs  couplings to electroweak gauge bosons are known 
to a precision between  ten and twenty percent
and Higgs  couplings to third generation fermions to about hundred percent. The Higgs 
boson couplings to first and second generation fermions 
are practically unconstrained. 
It is expected that the situation will dramatically improve with the continued operation of the LHC. 
For example, it is estimated \cite{CMS:2013xfa} that, by the end of the high-luminosity phase, the Higgs couplings, 
that can be extracted from its major production and decay  
channels,   will  be determined with a few percent precision.  There are several unknowns 
that may affect the validity of these projections,  including progress in reducing the 
uncertainties in theoretical predictions and the ability of experimentalists  to come up with 
new ideas but,  barring revolutionary breakthroughs, these estimates give us a ballpark 
of what can be expected.

Determination of Higgs couplings at the LHC 
requires  theoretical predictions for relevant processes, including both 
signal and background.  A case in point is the Higgs boson transverse momentum distribution, 
whose theoretical 
understanding is important to  properly  describe   the kinematics  of the Higgs decay products,  
but may also give us access to physics beyond the Standard Model~\cite{Grazzini:2016paz}.

Higgs bosons at the LHC are mostly produced in gluon collisions.  If additional 
gluons are radiated, a  Higgs boson recoils against them; this mechanisms  
leads   to a non-trivial Higgs $p_\perp$-spectrum whose theoretical 
description  requires good understanding of QCD dynamics. 
In the case of a point-like coupling of the Higgs to gluons,
%If the Higgs coupling to gluons was point-like,   
perturbative QCD (pQCD) provides an established  framework to describe the Higgs 
$p_\perp$-spectrum, including 
 fixed order QCD computations   recently extended to 
next-to-next-to-leading order  (NNLO) 
\cite{nnlohjet}  and the resummation  computations known in the 
next-to-next-to-leading logarithmic (NNLL) approximation 
%\cite{Bozzi:2003jy,Bozzi:2005wk,Monni:2016ktx}
\cite{resum1}.\footnote{ A related 
topic of  jet-veto resummation in Higgs production is discussed 
in  Refs.~\cite{resum2}
%~\cite{Stewart:2013faa,Becher:2013xia,Banfi:2012jm,Becher:2012qa,Tackmann:2012bt}.
}
However, the Higgs coupling to gluons in the 
Standard Model is the result of a quantum process where gluons  fluctuate into a quark-antiquark 
pair that annihilates into a Higgs boson.  
Because of the   differences in  fermion Yukawa couplings, 
the largest contribution to the $ggH$ coupling in the Standard Model comes 
from top quark loops, followed by contributions of   bottom and  charm quarks.
For values of the Higgs transverse momentum $p_\perp \ll m_t$, the top loop 
contribution to the $ggH$ coupling can be considered point-like to a very good approximation
and we can apply the full 
power of pQCD to describe it with high precision.
However, the bottom and charm loops are not point-like for moderate values 
of the transverse momentum 
and both the perturbative behavior and 
the possibility to perform resummations are much less understood for 
these contributions to the effective $ggH$ coupling.

Moreover, it is known that the bottom and charm  quark  contributions to $gg \to Hg $ amplitudes  
develop a peculiar, Sudakov-like dependence  on the Higgs boson transverse momentum \cite{Ellis:1987xu,bcontrib}. Taking 
the bottom quark contribution as an example, we find  
$A_{gg \to Hg}^{b}  \sim  m_b^2/m_H^2  \log^2 (p_\perp^2/m_b^2)$~\cite{Banfi:2013eda}. These double logarithms are not accounted 
for in the standard resummation framework\footnote{See Refs.\cite{logs}
%\cite{Melnikov:2016emg,Caola:2016upw} 
for recent attempts to understand the origin of these logarithms and 
the possibility to resum them. }  \cite{Grazzini:2013mca} and they significantly 
enhance the contribution of bottom loops to the Higgs production cross section in gluon 
fusion, compared to naive expectations.  In fact, the bottom loop contribution 
to Higgs production in the Standard Model is estimated to be close to minus five percent~\cite{Anastasiou:2016cez}
and, therefore, significant on the scale of  ${\cal O}(1\%)$ precision  goal discussed 
earlier. 

It is interesting to remark  that the ``substructure'' of the $ggH$ coupling 
is precisely what makes the Higgs transverse momentum distribution an interesting observable 
from the point of view of physics beyond the Standard Model. 
For example, current constraints on the  charm Yukawa coupling  are weak but, 
if the charm Yukawa coupling deviates significantly from its  Standard Model value, 
the charm contribution to $gg \to H$ increases,  and the relevance of the 
$c \bar c \to H$ annihilation channel for Higgs production grows. 
These modifications may result in observable effects in the Higgs transverse momentum distribution.
It was pointed out in Ref.~\cite{Bishara:2016jga} that studies of the Higgs boson transverse 
momentum distribution 
lead to  very competitive  constraints on the charm  Yukawa coupling; 
for example, 
it is expected~\cite{Bishara:2016jga} that 
at high-luminosity LHC, the charm Yukawa coupling  can be 
constrained to lie in the interval  $y_c/y_{c}^{\rm SM} \in [-2.9, 4.2] $ 
at the $95 \% $ confidence level.  Although not quite relevant for this paper, we also note 
that at very high values of the transverse momentum $p_\perp \gg m_t$,  the contribution 
of top quark loops can be resolved; this allows to probe for a  point-like 
component of the $gg H$ coupling that may originate from physics beyond the Standard Model.

This  discussion suggests that the shape of the Higgs boson 
transverse momentum distribution,  from moderate to high  $p_\perp$-values, 
is important for a proper description of the kinematic features 
of Higgs bosons produced at the LHC  and, also,  
may  provide  important information about physics beyond the Standard Model.  
Accurate Standard Model predictions for this observable are key for achieving 
these goals.  
As we already mentioned, the  pQCD description of  the Higgs boson 
transverse momentum distribution,  in the approximation 
of the point-like $ggH$ coupling,  is rather advanced, 
see Refs.~\cite{nnlohjet,Monni:2016ktx}, but there is very little 
understanding of how its not-point-like component is affected  by QCD 
radiative corrections.  To clarify this issue, we report on the computation 
of  QCD 
radiative corrections to top-bottom interference contribution to 
Higgs boson production at the LHC  in this Letter.

The calculation of the NLO QCD corrections to the top-bottom interference is non-trivial and we briefly 
summarize its salient details. The leading order production of the Higgs boson with non-vanishing 
transverse momentum occurs in different partonic channels, namely   $gg \to Hg$, $q g \to Hq $, 
$\bar q g \to H \bar q $ and $q \bar q \to H g$. At leading order these processes are mediated by 
top or bottom loops (the charm contribution in the SM is negligible). %that connect Higgs bosons to gluons. 
The one-loop  amplitudes are known exactly as functions of external kinematic variables and 
the quark masses \cite{Ellis:1987xu}.

At NLO, the production cross section  
receives contributions from real and virtual  corrections.  
Since the leading order process only occurs at one-loop, the virtual corrections 
require two-loop computations that include planar and non-planar box diagrams with internal masses. 
The computation of such Feynman diagrams is a matter of active current research that 
includes attempts 
to develop efficient numerical methods that can be used  in physical kinematics 
%\cite{Baernreuther:2013caa,Borowka:2016ypz,Borowka:2016ehy}, 
\cite{nummass}
and to extend  existing  analytic methods to make them applicable to 
two-loop Feynman diagrams with internal masses \cite{masses}. 

However, if we  focus on the  top-bottom interference and its impact on Higgs production at the LHC,  
we can simplify the calculation by  
using  the fact that  the mass of the $b$-quark, $m_b \sim 4.7~{\rm GeV}$,  is numerically small. 
Indeed, since $m_b  \ll m_H, p_\perp^{\rm typ}$, 
where $p_\perp^{\rm typ} \sim 30~{\rm GeV}$ is a typical Higgs boson transverse momentum, 
Feynman diagrams 
that  describe  Higgs production  can be expanded in series in $m_b$ for the purposes 
of LHC phenomenology.  We have checked at leading order that 
the use of scattering amplitudes either exact or expanded in $m_b$ 
leads to {\it at most} 
few  percent differences in the interference {\it contribution} 
to the Higgs $p_\perp$ distribution,  down to $p_\perp \sim 10~{\rm GeV}$. Since
the interference contribution changes the Higgs boson transverse momentum spectrum by ${\cal O}(5\%)$
at leading order,  the percent  difference between 
expanded and not expanded results is irrelevant  for phenomenology.

Unfortunately, the  expansion in $m_b$ is non-trivial since the Higgs boson production 
cross section  depends logarithmically on the $b$-quark mass. 
Therefore, we need to devise a procedure to expand scattering 
amplitudes in $m_b$ and  extract the non-analytic terms. 
This can be done
by deriving  differential  equations for master integrals 
that are needed to describe the two-loop corrections to $pp \to H+j$   
and then solving them in the limit $m_b \to 0$~\cite{Melnikov:2016qoc}. 
Indeed, since we can derive 
differential equations  to describe the dependence of the 
master integrals on the mass parameter $m_b$ and on the Mandelstam kinematic variables,  
and since all the information about singular points of a particular Feynman integral  
is contained in the differential equations that this Feynman integral satisfies, 
we can systematically solve the differential equation in series of $m_b$ and extract the 
non-analytic behavior.  
We note that a similar method was used to compute the 
top-bottom interference contribution to the  inclusive Higgs production cross section
in Ref.~\cite{Mueller:2015lrx}.

We have used this method  to calculate all the relevant 
two-loop scattering amplitudes to describe 
the production of a Higgs boson in association with a jet \cite{Melnikov:2016qoc,Melnikov:2017pgf}.
In our computation, all quarks in the initial and final 
states are massless, so that 
$b$-initiated processes are not included.   The two-loop amplitudes mediated by top quark loops, required 
to describe  the interference,  are computed in the approximation of an infinitely 
heavy top quark \cite{schmidt}. 

To produce physical results for  $H+j$ production, we need to combine 
the virtual corrections discussed  above with  the real corrections that describe 
inelastic processes, e.g. $gg \to H+gg$, $qg \to Hq+g$ etc. 
Computation of one-loop scattering amplitudes for these inelastic processes is non-trivial; 
it requires the evaluation of five-point Feynman integrals  with massive internal particles.
Nevertheless, such  amplitudes are known analytically since long ago~\cite{DelDuca} 
and were  recently  re-evaluated in Ref.~\cite{Neumann:2016dny}.

In this Letter we follow a different approach, based on the 
automated numerical computation of one-loop scattering amplitudes developed in recent years.  
One such approach, known as OpenLoops~\cite{Cascioli:2011va}, employs a hybrid tree-loop recursion.
Its implementation is publicly available~\cite{openloops} and
has been applied to compute one-loop QCD and electroweak corrections to
multi-leg scattering amplitudes for a variety of 
complicated processes  
(see e.g. Refs. \cite{olrefs,Jezo:2016ujg})  and as an input for the real-virtual 
contributions in NNLO computations (see e.g. Ref.~\cite{Cascioli:2014yka}).

For applications in NNLO calculations the corresponding real-virtual
one-loop contributions need to be computed in kinematic regions where one of the external partons becomes soft or collinear to other partons. We face a similar situation for the loop-induced process discussed in this Letter.
Indeed, the loop-squared real contribution has to be evaluated in phase-space regions where a final-state 
parton becomes
unresolved.
Although the singular contribution of the real emission graphs is easily identified and subtracted,  
it is important to control the approach of the singular region of the squared one-loop amplitudes. 
A reliable computation in such
kinematic regions is non-trivial, but 
the {\tt OpenLoops} approach appears to be  perfectly capable of coping with this challenge
thanks to the numerical stability of the employed algorithms. An important 
element of this stability is the program {\tt COLLIER}~\cite{collier} that is used 
to perform the tensor integral reduction in a clever way via expansions in small Gram determinants. 

We have implemented all 
virtual and real amplitudes in the {\tt POWHEG-BOX}~\cite{powheg}, where 
infra-red  singularities are regularized via FKS subtraction~\cite{Frixione:1995ms}. All {\tt OpenLoops} 
amplitudes 
are accessible via  a process-independent  interface developed in Ref.~\cite{Jezo:2016ujg}. The implementation
within the {\tt POWHEG-BOX} will allow for an easy  matching of the fixed-order results presented here with
parton showers at NLO. 
At leading order this has been done in Ref.~\cite{mctb}.

Using the methods described above, we calculated  the NLO QCD corrections to the top-bottom 
interference contribution to $H+j$ production in hadron collisions.
We identify the interference  contribution through 
its dependence on top-bottom Yukawa couplings. For the Higgs production 
cross section, we write 
\be
{\rm d} \sigma= {\rm d} \sigma_{tt} + {\rm d} \sigma_{tb} + {\rm d} \sigma_{bb},
\label{eq1}
\ee
where individual contributions to the differential cross section scale as 
${\rm d} \sigma_{tt} \sim {\cal O}(y_t^2)$,  
${\rm d} \sigma_{tb} \sim {\cal O}(y_t y_b)$, 
${\rm d} \sigma_{bb} \sim {\cal O}(y_b^2)$.
Given the hierarchy of the Yukawa couplings, $y_t \sim 1 \gg y_b \sim 10^{-2}$, 
the last term in Eq.(\ref{eq1})  can be safely neglected. Note, however, that if one focuses 
on Higgs-related observables that are inclusive with respect to the QCD radiation, 
${\rm d} \sigma_{bb}$ receives contributions from Higgs boson production in 
association with $b$-quarks, 
e.g.  $gg \to Hbb$. 
These processes change inclusive Higgs boson observables at below a permille level 
which makes them irrelevant unless b-jets in the final state are tagged.

Our main focus is the top-bottom interference contribution  ${\rm d} \sigma_{tb}$. 
Considering the virtual corrections, 
we write 
\be
{\rm d} \sigma_{tb}^{\rm virt}  \sim {\rm Re} \left[ A_t^{\rm LO}  A_b^{{\rm LO}*} + 
\frac{\alpha_s}{2 \pi}  ( A_t^{\rm NLO} A_b^{{\rm LO} *} + A_t^{\rm LO} A_b^{{\rm NLO}*} ) \right].
\label{eq2}
\ee
The leading order (one-loop)  term  in this formula is  known,   including  full 
 mass dependence. The NLO (two-loop) amplitudes  with the top quark $A_t^{\rm NLO}$ are only known 
in the limit $m_t \to \infty$ and we use $A_t^{\rm NLO}(m_t \to \infty)$ 
as an approximation for  $A_t^{\rm NLO}(m_t) $. In principle, one can improve on this by computing 
$1/m_t$ corrections to $A_t^{\rm NLO}(m_t \to \infty)$, see  Ref.~\cite{Harlander:2012hf},  
but it is not expected that 
such power corrections will have significant  impact on the  results for the interference 
at moderate, $p_\perp < m_t $,  values of the Higgs transverse momentum. The real emission contributions are 
computed with exact top- and bottom-mass dependence throughout the paper.

In what follows, we present the QCD corrections to  the top-bottom interference contribution 
to  the Higgs boson transverse momentum distribution and to the 
Higgs rapidity distribution in $H+j$ production.   We consider proton collisions at the $13~{\rm TeV}$ LHC and 
take the  mass of the Higgs boson  to be   $m_H = 125~{\rm GeV}$.  

We work within a fixed flavor-number scheme and do not consider bottom quarks as partons in 
the proton.  We use the NNPDF30 set of parton distribution functions \cite{Ball:2014uwa}. We also 
use the strong 
coupling constant $\alpha_s(M_Z)$ 
that is provided with this  PDF set. We renormalize 
the $b$-quark mass in the on-shell scheme and use $m_b~=~4.75~{\rm GeV}$ as its numerical value. 
We choose renormalization and factorization scales to be equal and take, 
as the central value
$
%\mu = H_T/2, \;\;\; H_T = \sqrt{m_H^2 + p_{H,\perp}^2} + \sum_{j}^{}  p_{T,j}, 
\mu = H_T/2, \;\;\; H_T = \sqrt{m_H^2 + p_{\perp}^2} + \sum_{j}^{}  p_{\perp,j},
$
where the sum runs over all partons in the final state. 

To quantify the impact  of the top-bottom interference on an 
observable ${\cal O}$, it is convenient to define the following quantity 
\be
{\cal R}_{\rm int}\left [ {\cal O} \right ]   = \frac{ \int {\rm d} \sigma_{tb} \; \delta({\cal O} - {\cal O}(\vec x))  }{ 
\int {\rm d} \sigma_{tt} \; \delta({\cal O} - {\cal O}(\vec x))  },
\label{eq1}
\ee
where  $\vec x$ is a set of phase-space variables. 
  Note that we do not expand 
the $\sigma_{tt}$ cross section in the denominator in Eq.(\ref{eq1}) in powers of $\alpha_s$.  
Therefore, any change in  ${\cal R}_{\rm int}$ in consecutive orders in perturbation theory 
would reflect  {\it differences} 
in QCD corrections to the $tb$ interference and the point-like contribution  to $H+j$ production.   
In what follows we present ${\cal R}_{\rm int}$ as a function of the 
Higgs boson transverse momentum $p_\perp$ and the (pseudo-)rapidity $\eta_H$.

\begin{figure}[t]
\centering
\includegraphics[width=0.5\textwidth]{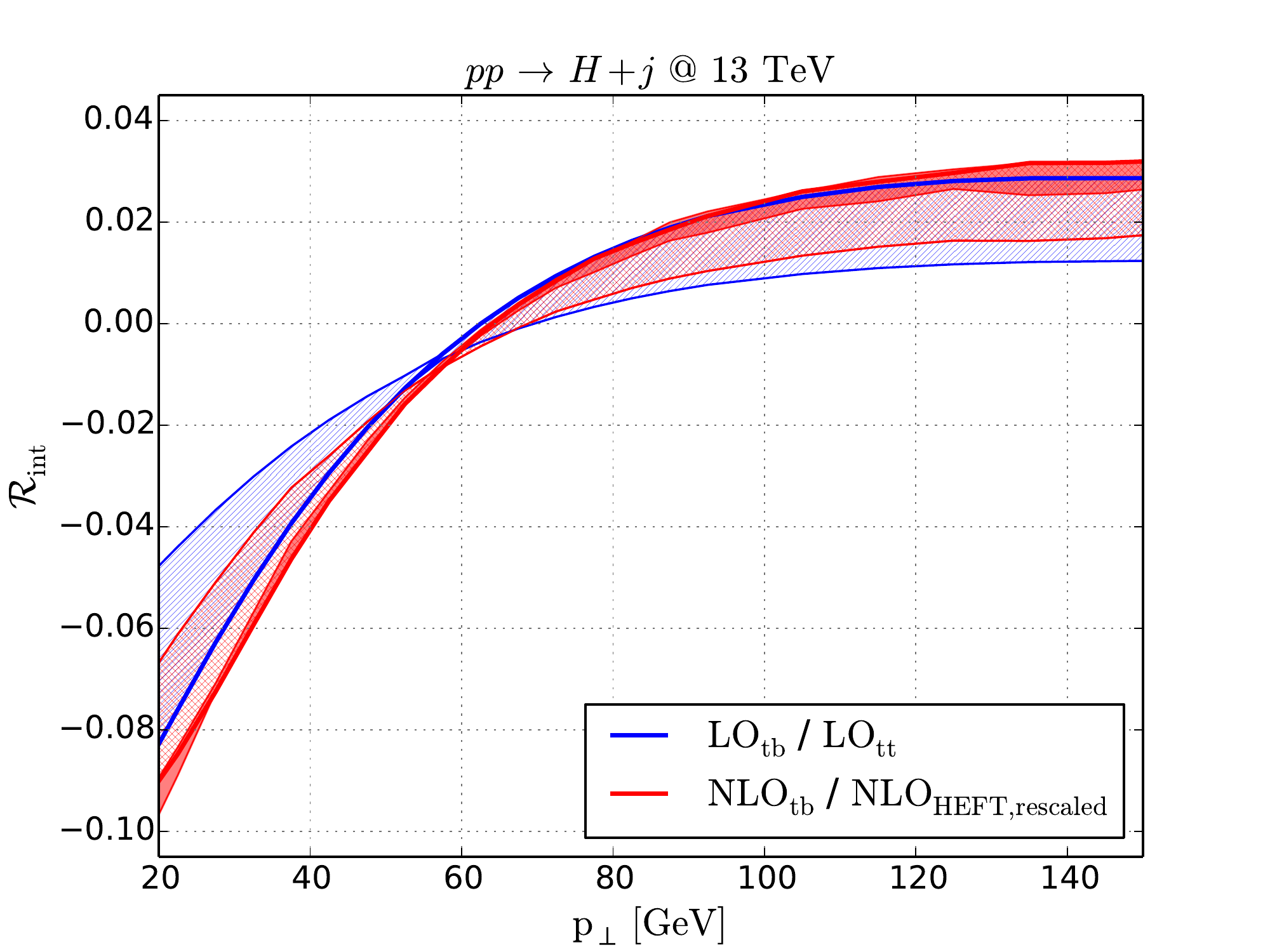}
\caption{Relative top-bottom interference contribution to the transverse momentum distribution of the Higgs boson at leading (blue) and next-to-leading (red) order in perturbative QCD. At next-to-leading order the interference contribution is shown with respect to the point-like Higgs Effective Field Theory prediction rescaled with exact leading-order top mass dependence.
Filled bands, hardly visible at leading order, show the change in ${\cal R}_{\rm int}$ caused by a variation of the renormalization and factorization scales, correlated between numerator and denominator. The hashed bands indicate the uncertainty due to mass-renormalization scheme variation. See text for details. 
}\label{fig2}
\end{figure}

\begin{figure}[t]
\centering
\includegraphics[width=0.5\textwidth]{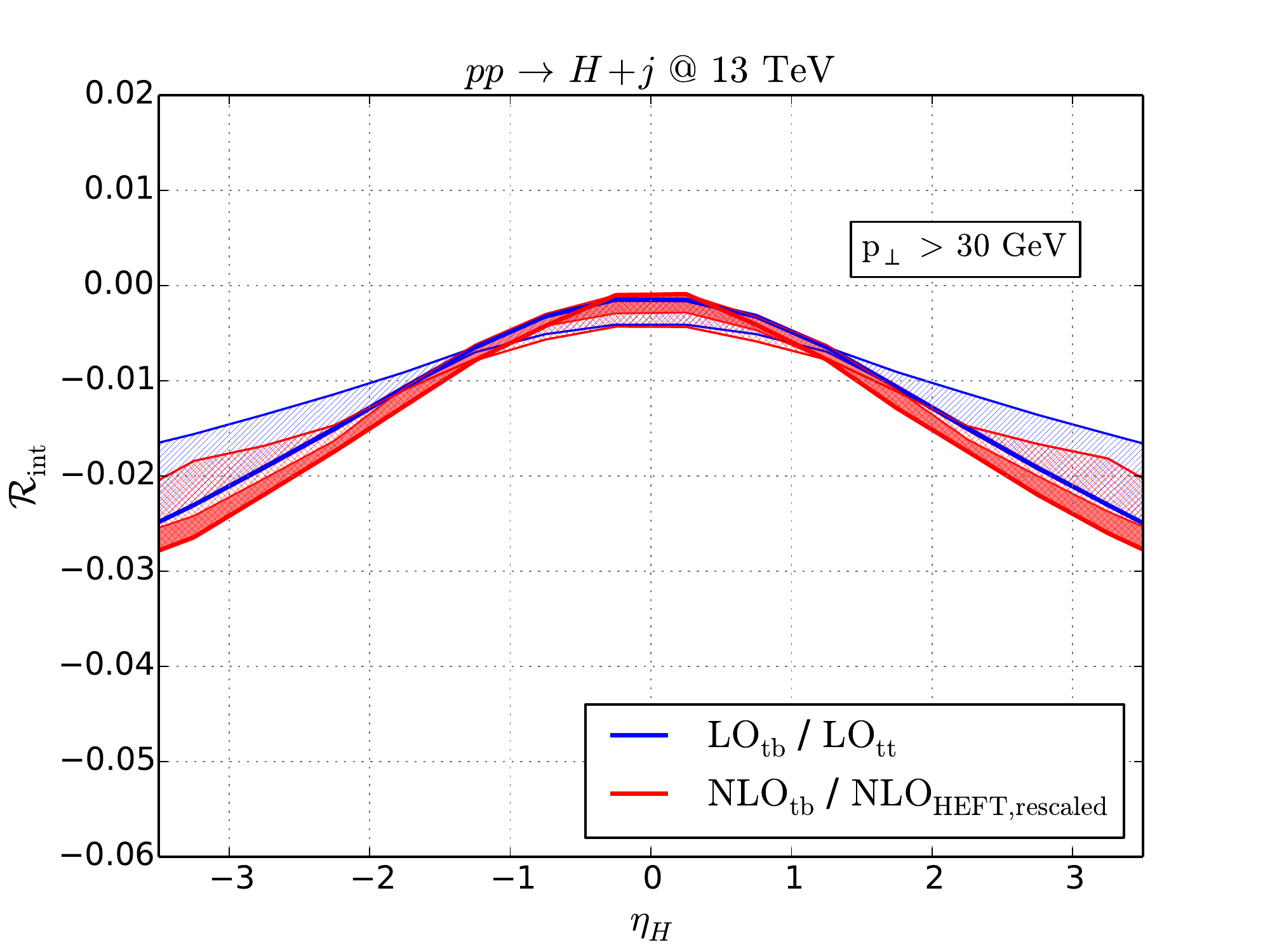}
\caption{ Relative top-bottom interference contribution to the pseudo-rapidity distribution of the Higgs boson at leading and next-to-leading 
order in perturbative QCD. Bands and colors as in Fig.\ref{fig2}.
}\label{fig1}
\end{figure}

The impact of the top-bottom interference  on the Higgs boson 
transverse momentum distribution is shown in Fig.~\ref{fig2}.  We observe that the leading 
order interference changes the Higgs boson transverse momentum distribution by 
$-8\%$ at $p_\perp \sim 20~{\rm GeV}$ and $+2\%$ at $p_\perp \sim 100~{\rm GeV}$. 
Since the QCD corrections  to color-singlet production in gluon annihilation are large 
and since it is not clear a priori if the QCD corrections to the interference are similar 
to the QCD corrections to the point-like cross section, 
large modifications of these LO results can not be excluded.   The NLO computation, illustrated 
in Fig.~\ref{fig2},  clarifies this point.  
There, filled bands in blue for the leading and red for the next-to-leading order predictions show the result 
for ${\cal R}_{\rm int}(p_{\perp})$ 
computed in the pole mass renormalization scheme.
The widths of the bands indicate changes in the predictions caused by 
variations of renormalization and factorization scales by a factor of 
two around the central value $\mu=H_{T}/2$. In fact, we observe that differences between leading and next-to-leading order are very small. For example, 
${\cal R}^{\rm NLO}_{\rm int}(p_\perp)$ appears to be smaller than 
${\cal R}^{\rm LO}_{\rm int}(p_\perp)$ by less than a percent at $p_\perp < 60~{\rm GeV}$ and, practically, 
coincides with it at higher values of $p_\perp$. 
We emphasise that these small changes in ${\cal R}_{\rm int}$ imply \emph{sizable}, ${\cal O}(40-50\%)$, corrections
to the $tb$ interference proper
that,  however, appear to be very similar to NLO QCD corrections to the point-like cross section $\sigma_{tt}$. 
The scale variation bands are very narrow (at leading-order hardly visible) due to a cancellation of large scale variation changes between numerator and denominator in Eq.(\ref{eq1}).   Similar results for the Higgs boson rapidity 
distribution for events with $p_\perp > 30~{\rm GeV}$ are shown in Fig.~\ref{fig1}.

The above result for the scale variation suggests that  the uncertainties in predicting 
the size of top-bottom interference effects in $H+j$ production are small 
since both the size 
of corrections and the scale variation bands are similar to the corrections to the point-like $pp \to H+j$ 
cross section.  Such a conclusion, nevertheless, misses an important  source of uncertainties related 
to  a possible choice of  a different mass-renormalization scheme. 
Indeed, since the leading order interference contribution is proportional to the square of the bottom 
mass  $R_{\rm int} \sim m_b^2$ and since at leading order a change in the mass renormalization scheme simply 
amounts to the use of different numerical values for $m_b$ in calculating  ${\cal R}_{\rm int}$, it is 
easy to see that this ambiguity is very significant. Indeed, suppose that we choose to renormalize the bottom
mass in the $\overline {\rm MS}$ scheme and we take $m_b = m_b^{\overline {\rm MS}}(100~{\rm GeV})~=~3.07~{\rm GeV}$ as 
input parameter.\footnote{We calculated this value using the program RunDec \cite{rundec} 
with the input value $ m_b^{\overline{ \rm MS}} ( m_b^{\overline{ \rm MS}} )~=~4.2~{\rm GeV}$.}
 Since $ ( m_b^{\overline {\rm MS}}(100~{\rm GeV})/m_b^{\rm pole} )^2~\approx~0.4$,  
this implies that ${\cal R}_{\rm int}^{\rm LO}$ is  reduced by more than a factor of two, practically 
independent 
of the $p_\perp$ value.  This large leading order variation is shown as a hashed blue band in Figs.~\ref{fig2},\ref{fig1}, 
where we have taken $m_b = m_b^{\rm pole}$ and $m_b = m_b^{\overline {\rm MS}}(100~{\rm GeV})$ 
as the two boundary values. 

This large ambiguity in the leading order value of ${\cal R}_{\rm int}$  is somewhat reduced 
at next-to-leading order where the effect of the mass renormalization 
scheme change is less dramatic but, nevertheless significant. 
The scheme dependence at NLO, for the setup explained in the previous paragraph, is shown as a hashed red band. 
We observe that for $p_\perp < 60~{\rm GeV}$, the mass renormalization scheme uncertainty is reduced by almost 
a factor of two, whereas the reduction of uncertainty is only marginal at higher $p_\perp$.  
This happens because 
in the final result for the interference at high transverse momenta 
there is a significant cancellation between 
$A_t^{\rm NLO} A_b^{\rm LO *}$ and  $A_t^{\rm LO} A_b^{\rm NLO *}$, c.f. Eq.(\ref{eq2}).  Since the first term 
involves leading order $b$-quark contributions, it experiences large variations when the $b$-quark 
mass renormalization scheme is changed and this causes large variations in ${\cal R}^{\rm int}$ 
at high $p_\perp$.  The interference contribution to the Higgs rapidity distribution in Fig.~\ref{fig1} 
shows similar features. The mass variation band at NLO is smaller than the LO variation band at 
large absolute values of the pseudo-rapidity (dominated by small $p_\perp$) and practically indistinguishable from it 
at the central rapidity values (dominated by large $p_\perp$). 

In summary, we computed the NLO QCD corrections to the top-bottom interference contribution to 
Higgs boson production in association with a jet at the LHC. This is the first computation 
of QCD radiative corrections to  Higgs production at this order in perturbation theory that 
goes beyond the point-like approximation for the $ggH$ coupling. Our results show that corrections 
to the interference are large yet they appear to track very well corrections to the point-like 
component of the cross section.  The strong dependence of the LO interference 
on the mass-renormalization scheme is reduced at NLO but  at high values of 
the Higgs transverse momentum 
or at central rapidity, the remaining ambiguities are  significant. 
It is not clear how the situation 
at high $p_\perp$ and/or small absolute $\eta_H$ 
can be further improved. However, we want to emphasize that in these kinematic 
regions the interference is numerically small compared to the ${\cal O}(y_t^2)$ contribution. 
Nevertheless, with this result at hand, one can try to 
provide the best possible theoretical predictions for the Higgs transverse momentum distribution that combine 
the known results for the $p_\perp$-resummation, NNLO corrections to $H+j$ in the point-like 
approximation  with the top-bottom interference.
All the ingredients are now available. 
We plan to return to this problem  before 
long.

{\bf Acknowledgments}
We thank Tomas Jezo for valuable help with the {\tt POWHEG-BOX} and
Fabrizio Caola for useful conversations.  The
research of K.M. was supported by the German Federal Ministry for 
Education and Research (BMBF) under
grant 05H15VKCCA.


\begin{thebibliography}{99}


%\cite{Gupta:2013zza}
\bibitem{Gupta:2013zza} 
  R.~S.~Gupta, H.~Rzehak and J.~D.~Wells,
  %``How well do we need to measure the Higgs boson mass and self-coupling?,''
  Phys.\ Rev.\ D {\bf 88}, 055024 (2013).


%\cite{Khachatryan:2016vau}
\bibitem{Khachatryan:2016vau} 
  G.~Aad {\it et al.} [ATLAS and CMS Collaborations],
  %``Measurements of the Higgs boson production and decay rates and constraints on its couplings from a combined ATLAS and CMS analysis of the LHC pp collision data at $ \sqrt{s}=7 $ and 8 TeV,''
  JHEP {\bf 1608}, 045 (2016).

\bibitem{CMS:2013xfa} 
  [CMS Collaboration],
  ``Projected Performance of an Upgraded CMS Detector at the LHC and HL-LHC: Contribution to the Snowmass Process,''
  arXiv:1307.7135 [hep-ex].
  %%CITATION = ARXIV:1307.7135;%%

\bibitem{Grazzini:2016paz} 
  M.~Grazzini, A.~Ilnicka, M.~Spira and M.~Wiesemann,
  %``Modeling BSM effects on the Higgs transverse-momentum spectrum in an EFT approach,''
  arXiv:1612.00283 [hep-ph].

\bibitem{Bishara:2016jga} 
  F.~Bishara, U.~Haisch, P.~F.~Monni and E.~Re,
  %``Constraining Light-Quark Yukawa Couplings from Higgs Distributions,''
  arXiv:1606.09253 [hep-ph].
  %%CITATION = ARXIV:1606.09253;%%


\bibitem{nnlohjet}
X.~Chen, T.~Gehrmann, E.~W.~N.~Glover and M.~Jaquier,
  %``Precise QCD predictions for the production of Higgs + jet final states,''
  Phys.\ Lett.\ B {\bf 740} (2015) 147;
R.~Boughezal, F.~Caola, K.~Melnikov, F.~Petriello and M.~Schulze,
  %``Higgs boson production in association with a jet at next-to-next-to-leading order,''
  Phys.\ Rev.\ Lett.\  {\bf 115} (2015) no.8,  082003;
F.~Caola, K.~Melnikov and M.~Schulze,
  %``Fiducial cross sections for Higgs boson production in association with a jet at next-to-next-to-leading order in QCD,''
  Phys.\ Rev.\ D {\bf 92} (2015) no.7,  074032;
X.~Chen, J.~Cruz-Martinez, T.~Gehrmann, E.~W.~N.~Glover and M.~Jaquier,
  %``NNLO QCD corrections to Higgs boson production at large transverse momentum,''
  JHEP {\bf 1610} (2016) 066;
R.~Boughezal, C.~Focke, W.~Giele, X.~Liu and F.~Petriello,
  %``Higgs boson production in association with a jet at NNLO using jettiness subtraction,''
  Phys.\ Lett.\ B {\bf 748} (2015) 5.




%%\cite{Bozzi:2003jy}
%\bibitem{Bozzi:2003jy} 
%  G.~Bozzi, S.~Catani, D.~de Florian and M.~Grazzini,
%  %``The q(T) spectrum of the Higgs boson at the LHC in QCD perturbation theory,''
%  Phys.\ Lett.\ B {\bf 564}, 65 (2003).
%  %%CITATION = doi:10.1016/S0370-2693(03)00656-7;%%
%  %224 citations counted in INSPIRE as of 16 Feb 2017
%
%
%
%%\cite{Bozzi:2005wk}
%\bibitem{Bozzi:2005wk} 
%  G.~Bozzi, S.~Catani, D.~de Florian and M.~Grazzini,
%  %``Transverse-momentum resummation and the spectrum of the Higgs boson at the LHC,''
%  Nucl.\ Phys.\ B {\bf 737}, 73 (2006).
%  %%CITATION = doi:10.1016/j.nuclphysb.2005.12.022;%%
%  %333 citations counted in INSPIRE as of 16 Feb 2017
%
%
%%\cite{Monni:2016ktx}
\bibitem{Monni:2016ktx} 
  P.~F.~Monni, E.~Re and P.~Torrielli,
  %``Higgs Transverse-Momentum Resummation in Direct Space,''
  Phys.\ Rev.\ Lett.\  {\bf 116}, no. 24, 242001 (2016).
%  doi:10.1103/PhysRevLett.116.242001
%  [arXiv:1604.02191 [hep-ph]].
  %%CITATION = doi:10.1103/PhysRevLett.116.242001;%%

\bibitem{resum1}
  G.~Bozzi, S.~Catani, D.~de Florian and M.~Grazzini,
  %``The q(T) spectrum of the Higgs boson at the LHC in QCD perturbation theory,''
  Phys.\ Lett.\ B {\bf 564}, 65 (2003);
  %%CITATION = doi:10.1016/S0370-2693(03)00656-7;%%
  %224 citations counted in INSPIRE as of 16 Feb 2017
    G.~Bozzi, S.~Catani, D.~de Florian and M.~Grazzini,
  %``Transverse-momentum resummation and the spectrum of the Higgs boson at the LHC,''
  Nucl.\ Phys.\ B {\bf 737}, 73 (2006);
  %%CITATION = doi:10.1016/j.nuclphysb.2005.12.022;%%
  %333 citations counted in INSPIRE as of 16 Feb 2017
    P.~F.~Monni, E.~Re and P.~Torrielli,
  %``Higgs Transverse-Momentum Resummation in Direct Space,''
  Phys.\ Rev.\ Lett.\  {\bf 116}, no. 24, 242001 (2016).
%  doi:10.1103/PhysRevLett.116.242001
%  [arXiv:1604.02191 [hep-ph]].
  %%CITATION = doi:10.1103/PhysRevLett.116.242001;%%

\bibitem{resum2}
  I.~W.~Stewart, F.~J.~Tackmann, J.~R.~Walsh and S.~Zuberi,
  %``Jet $p_T$ resummation in Higgs production at $NNLL'+NNLO$,''
  Phys.\ Rev.\ D {\bf 89}, no. 5, 054001 (2014);
  %doi:10.1103/PhysRevD.89.054001
  %[arXiv:1307.1808 [hep-ph]].
  %%CITATION = doi:10.1103/PhysRevD.89.054001;%%
  T.~Becher, M.~Neubert and L.~Rothen,
  %``Factorization and $N^{3}LL_{p}$+NNLO predictions for the Higgs cross section with a jet veto,''
  JHEP {\bf 1310}, 125 (2013);
  %doi:10.1007/JHEP10(2013)125
  %[arXiv:1307.0025 [hep-ph]].
  %%CITATION = doi:10.1007/JHEP10(2013)125;%%
    A.~Banfi, P.~F.~Monni, G.~P.~Salam and G.~Zanderighi,
  %``Higgs and Z-boson production with a jet veto,''
  Phys.\ Rev.\ Lett.\  {\bf 109}, 202001 (2012);
  %doi:10.1103/PhysRevLett.109.202001
  %[arXiv:1206.4998 [hep-ph]].
  %%CITATION = doi:10.1103/PhysRevLett.109.202001;%%
    F.~J.~Tackmann, J.~R.~Walsh and S.~Zuberi,
  %``Resummation Properties of Jet Vetoes at the LHC,''
  Phys.\ Rev.\ D {\bf 86}, 053011 (2012);
  %doi:10.1103/PhysRevD.86.053011
  %[arXiv:1206.4312 [hep-ph]].
  %%CITATION = doi:10.1103/PhysRevD.86.053011;%%
    T.~Becher and M.~Neubert,
  %``Factorization and NNLL Resummation for Higgs Production with a Jet Veto,''
  JHEP {\bf 1207}, 108 (2012).
  %doi:10.1007/JHEP07(2012)108
  %[arXiv:1205.3806 [hep-ph]].
  %%CITATION = doi:10.1007/JHEP07(2012)108;%%

%\bibitem{Stewart:2013faa} 
%  I.~W.~Stewart, F.~J.~Tackmann, J.~R.~Walsh and S.~Zuberi,
%  %``Jet $p_T$ resummation in Higgs production at $NNLL'+NNLO$,''
%  Phys.\ Rev.\ D {\bf 89}, no. 5, 054001 (2014).
%  %doi:10.1103/PhysRevD.89.054001
%  %[arXiv:1307.1808 [hep-ph]].
%  %%CITATION = doi:10.1103/PhysRevD.89.054001;%%
%
%
%
%\bibitem{Becher:2013xia} 
%  T.~Becher, M.~Neubert and L.~Rothen,
%  %``Factorization and $N^{3}LL_{p}$+NNLO predictions for the Higgs cross section with a jet veto,''
%  JHEP {\bf 1310}, 125 (2013).
%  %doi:10.1007/JHEP10(2013)125
%  %[arXiv:1307.0025 [hep-ph]].
%  %%CITATION = doi:10.1007/JHEP10(2013)125;%%
%
%
%\bibitem{Banfi:2012jm} 
%  A.~Banfi, P.~F.~Monni, G.~P.~Salam and G.~Zanderighi,
%  %``Higgs and Z-boson production with a jet veto,''
%  Phys.\ Rev.\ Lett.\  {\bf 109}, 202001 (2012).
%  %doi:10.1103/PhysRevLett.109.202001
%  %[arXiv:1206.4998 [hep-ph]].
%  %%CITATION = doi:10.1103/PhysRevLett.109.202001;%%
%
%
%
%\bibitem{Tackmann:2012bt} 
%  F.~J.~Tackmann, J.~R.~Walsh and S.~Zuberi,
%  %``Resummation Properties of Jet Vetoes at the LHC,''
%  Phys.\ Rev.\ D {\bf 86}, 053011 (2012).
%  %doi:10.1103/PhysRevD.86.053011
%  %[arXiv:1206.4312 [hep-ph]].
%  %%CITATION = doi:10.1103/PhysRevD.86.053011;%%
%
%%\cite{Becher:2012qa}
%\bibitem{Becher:2012qa} 
%  T.~Becher and M.~Neubert,
%  %``Factorization and NNLL Resummation for Higgs Production with a Jet Veto,''
%  JHEP {\bf 1207}, 108 (2012).
%  %doi:10.1007/JHEP07(2012)108
%  %[arXiv:1205.3806 [hep-ph]].
%  %%CITATION = doi:10.1007/JHEP07(2012)108;%%


%\cite{Ellis:1987xu}
\bibitem{Ellis:1987xu}
  R.~K.~Ellis, I.~Hinchliffe, M.~Soldate and J.~J.~van der Bij,
  %``Higgs Decay to tau+ tau-: A Possible Signature of Intermediate Mass Higgs Bosons at the SSC,''
  Nucl.\ Phys.\ B {\bf 297}, 221 (1988).
%  doi:10.1016/0550-3213(88)90019-3
  %%CITATION = doi:10.1016/0550-3213(88)90019-3;%%


\bibitem{bcontrib}
  B.~Field, S.~Dawson and J.~Smith,
  %``Scalar and pseudoscalar Higgs boson plus one jet production at the CERN LHC and Tevatron,''
  Phys.\ Rev.\ D {\bf 69}, 074013 (2004);
%  doi:10.1103/PhysRevD.69.074013
  %[hep-ph/0311199].
  %%CITATION = doi:10.1103/PhysRevD.69.074013;%%
  %41 citations counted in INSPIRE as of 02 Mar 2017
  W.~Y.~Keung and F.~J.~Petriello,
  %``Electroweak and finite quark-mass effects on the Higgs boson transverse momentum distribution,''
  Phys.\ Rev.\ D {\bf 80}, 013007 (2009).
%  doi:10.1103/PhysRevD.80.013007
%  [arXiv:0905.2775 [hep-ph]].
  %%CITATION = doi:10.1103/PhysRevD.80.013007;%%
  O.~Brein,
  %``Electroweak and Bottom Quark Contributions to Higgs Boson plus Jet Production,''
  Phys.\ Rev.\ D {\bf 81}, 093006 (2010)
  

\bibitem{Banfi:2013eda} 
  A.~Banfi, P.~F.~Monni and G.~Zanderighi,
  %``Quark masses in Higgs production with a jet veto,''
  JHEP {\bf 1401}, 097 (2014)
  %doi:10.1007/JHEP01(2014)097
  %[arXiv:1308.4634 [hep-ph]].
  %%CITATION = doi:10.1007/JHEP01(2014)097;%%

\bibitem{logs}
  K.~Melnikov and A.~Penin,
  %``On the light quark mass effects in Higgs boson production in gluon fusion,''
  JHEP {\bf 1605}, 172 (2016);
%%CITATION = doi:10.1007/JHEP05(2016)172;%%
  F.~Caola, S.~Forte, S.~Marzani, C.~Muselli and G.~Vita,
  %``The Higgs transverse momentum spectrum with finite quark masses beyond leading order,''
  JHEP {\bf 1608}, 150 (2016).
%%CITATION = doi:10.1007/JHEP08(2016)150;%%

%\bibitem{Melnikov:2016emg} 
%  K.~Melnikov and A.~Penin,
%  %``On the light quark mass effects in Higgs boson production in gluon fusion,''
%  JHEP {\bf 1605}, 172 (2016).
%%%CITATION = doi:10.1007/JHEP05(2016)172;%%
%
%
%\bibitem{Caola:2016upw} 
%  F.~Caola, S.~Forte, S.~Marzani, C.~Muselli and G.~Vita,
%  %``The Higgs transverse momentum spectrum with finite quark masses beyond leading order,''
%  JHEP {\bf 1608}, 150 (2016).
%%%CITATION = doi:10.1007/JHEP08(2016)150;%%


\bibitem{Grazzini:2013mca} 
  M.~Grazzini and H.~Sargsyan,
  %``Heavy-quark mass effects in Higgs boson production at the LHC,''
  JHEP {\bf 1309}, 129 (2013).
%%CITATION = doi:10.1007/JHEP09(2013)129;%%


\bibitem{Anastasiou:2016cez} 
  C.~Anastasiou, C.~Duhr, F.~Dulat, E.~Furlan, T.~Gehrmann, F.~Herzog, A.~Lazopoulos and B.~Mistlberger,
  %``High precision determination of the gluon fusion Higgs boson cross-section at the LHC,''
  JHEP {\bf 1605}, 058 (2016)
  %doi:10.1007/JHEP05(2016)058
  %[arXiv:1602.00695 [hep-ph]].
  %%CITATION = doi:10.1007/JHEP05(2016)058;%%


\bibitem{mctb}
J.~Alwall, Q.~Li and F.~Maltoni,
  %``Matched predictions for Higgs production via heavy-quark loops in the SM and beyond,''
  Phys.\ Rev.\ D {\bf 85}, 014031 (2012)
E.~Bagnaschi, G.~Degrassi, P.~Slavich and A.~Vicini,
  %``Higgs production via gluon fusion in the POWHEG approach in the SM and in the MSSM,''
  JHEP {\bf 1202}, 088 (2012) 
R.~V.~Harlander, H.~Mantler and M.~Wiesemann,
 JHEP {\bf 1411}, 116 (2014);
M.~Buschmann, D.~Goncalves, S.~Kuttimalai, M.~Sch\"{o}nherr, F.~Krauss and T.~Plehn,
 JHEP {\bf 1502}, 038 (2015);
 K.~Hamilton, P.~Nason and G.~Zanderighi,
  %``Finite quark-mass effects in the NNLOPS POWHEG+MiNLO Higgs generator,''
  JHEP {\bf 1505}, 140 (2015)
E.~Bagnaschi and A.~Vicini,
 JHEP {\bf 1601}, 056 (2016);
E.~Bagnaschi, R.~V.~Harlander, H.~Mantler, A.~Vicini and M.~Wiesemann,
 JHEP {\bf 1601}, 090 (2016);
R.~Frederix, S.~Frixione, E.~Vryonidou and M.~Wiesemann,
  %``Heavy-quark mass effects in Higgs plus jets production,''
  JHEP {\bf 1608}, 006 (2016)
N.~Greiner, S.~H\"{o}che, G.~Luisoni, M.~Sch\"{o}nherr and J.~C.~Winter,
 JHEP {\bf 1701}, 091 (2017)


\bibitem{nummass}
  P.~B\"arnreuther, M.~Czakon and P.~Fiedler,
  %``Virtual amplitudes and threshold behaviour of hadronic top-quark pair-production cross sections,''
  JHEP {\bf 1402}, 078 (2014);
%%CITATION = doi:10.1007/JHEP02(2014)078;%%
  S.~Borowka, N.~Greiner, G.~Heinrich, S.~P.~Jones, M.~Kerner, J.~Schlenk and T.~Zirke,
  %``Full top quark mass dependence in Higgs boson pair production at NLO,''
  JHEP {\bf 1610}, 107 (2016);
%%CITATION = doi:10.1007/JHEP10(2016)107;%%
  S.~Borowka, N.~Greiner, G.~Heinrich, S.~P.~Jones, M.~Kerner, J.~Schlenk and T.~Zirke,
  %``Full top quark mass dependence in Higgs boson pair production at NLO,''
  JHEP {\bf 1610}, 107 (2016).
%%CITATION = doi:10.1007/JHEP10(2016)107;%%

%\bibitem{Baernreuther:2013caa} 
%  P.~B\"arnreuther, M.~Czakon and P.~Fiedler,
%  %``Virtual amplitudes and threshold behaviour of hadronic top-quark pair-production cross sections,''
%  JHEP {\bf 1402}, 078 (2014).
%%%CITATION = doi:10.1007/JHEP02(2014)078;%%
%
%\bibitem{Borowka:2016ypz} 
%  S.~Borowka, N.~Greiner, G.~Heinrich, S.~P.~Jones, M.~Kerner, J.~Schlenk and T.~Zirke,
%  %``Full top quark mass dependence in Higgs boson pair production at NLO,''
%  JHEP {\bf 1610}, 107 (2016).
%%%CITATION = doi:10.1007/JHEP10(2016)107;%%
%
%
%\bibitem{Borowka:2016ehy} 
%  S.~Borowka, N.~Greiner, G.~Heinrich, S.~P.~Jones, M.~Kerner, J.~Schlenk, 
%U.~Schubert and T.~Zirke,
%  %``Higgs Boson Pair Production in Gluon Fusion at Next-to-Leading Order with Full Top-Quark Mass Dependence,''
%  Phys.\ Rev.\ Lett.\  {\bf 117}, no. 1, 012001 (2016);
%  Erratum: {\it ibid}, {\bf 117}, no. 7, 079901 (2016).
%%%CITATION = doi:10.1103/PhysRevLett.117.079901, 10.1103/PhysRevLett.117.012001;%%




\bibitem{masses}
L.~Adams, C.~Bogner and S.~Weinzierl,
  %``The two-loop sunrise integral around four space-time dimensions and generalisations of the Clausen and Glaisher functions towards the elliptic case,''
  J.\ Math.\ Phys.\  {\bf 56}, no. 7, 072303 (2015);
  %doi:10.1063/1.4926985
  %[arXiv:1504.03255 [hep-ph]].
  %%CITATION = doi:10.1063/1.4926985;%%
  L.~Adams, C.~Bogner and S.~Weinzierl,
  %``The iterated structure of the all-order result for the two-loop sunrise integral,''
  J.\ Math.\ Phys.\  {\bf 57}, no. 3, 032304 (2016);
  %doi:10.1063/1.4944722
 % [arXiv:1512.05630 [hep-ph]].
  %%CITATION = doi:10.1063/1.4944722;%%
  E.~Remiddi and L.~Tancredi,
  %``Differential equations and dispersion relations for Feynman amplitudes. The two-loop massive sunrise and the kite integral,''
  Nucl.\ Phys.\ B {\bf 907} 400 (2016);
  %doi:10.1016/j.nuclphysb.2016.04.013
  %[arXiv:1602.01481 [hep-ph]].
  %%CITATION = doi:10.1016/j.nuclphysb.2016.04.013;%%
   L.~Adams, C.~Bogner, A.~Schweitzer and S.~Weinzierl,
  %``The kite integral to all orders in terms of elliptic polylogarithms,''
  %doi:10.1063/1.4969060
  arXiv:1607.01571 [hep-ph];
  %%CITATION = doi:10.1063/1.4969060;%%
  R.~Bonciani, V.~Del Duca, H.~Frellesvig, J.~M.~Henn, F.~Moriello and V.~A.~Smirnov,
  %``Two-loop planar master integrals for Higgs$\to 3$ partons with full heavy-quark mass dependence,''
  JHEP {\bf 1612}, 096 (2016);
%  doi:10.1007/JHEP12(2016)096
%  [arXiv:1609.06685 [hep-ph]].
  %%CITATION = doi:10.1007/JHEP12(2016)096;%%
  A.~Primo and L.~Tancredi,
  %``On the maximal cut of Feynman integrals and the solution of their differential equations,''
  Nucl.\ Phys.\ B {\bf 916}, 94 (2017);
  %doi:10.1016/j.nuclphysb.2016.12.021
  %[arXiv:1610.08397 [hep-ph]].
  %%CITATION = doi:10.1016/j.nuclphysb.2016.12.021;%%
   A.~von Manteuffel and L.~Tancredi,
  %``A non-planar two-loop three-point function beyond multiple polylogarithms,''
  arXiv:1701.05905 [hep-ph];
  %%CITATION = ARXIV:1701.05905;%%
  H.~Frellesvig and C.~G.~Papadopoulos,
  %``Cuts of Feynman Integrals in Baikov representation,''
  arXiv:1701.07356 [hep-ph].
  %%CITATION = ARXIV:1701.07356;%%

%\cite{Melnikov:2016qoc}
\bibitem{Melnikov:2016qoc} 
  K.~Melnikov, L.~Tancredi and C.~Wever,
  %``Two-loop $gg \to Hg$ amplitude mediated by a nearly massless quark,''
  JHEP {\bf 1611}, 104 (2016).
%  doi:10.1007/JHEP11(2016)104
%  [arXiv:1610.03747 [hep-ph]].
  %%CITATION = doi:10.1007/JHEP11(2016)104;%%


%\cite{Mueller:2015lrx}
\bibitem{Mueller:2015lrx} 
  R.~Mueller and D.~G.~\"Ozt\"{u}rk,
  %``On the computation of finite bottom-quark mass effects in Higgs boson production,''
  JHEP {\bf 1608}, 055 (2016).
%  doi:10.1007/JHEP08(2016)055
%  [arXiv:1512.08570 [hep-ph]].
  %%CITATION = doi:10.1007/JHEP08(2016)055;%%

%\cite{Melnikov:2017pgf}
\bibitem{Melnikov:2017pgf} 
  K.~Melnikov, L.~Tancredi and C.~Wever,
  %``Two-loop amplitudes for $q g \to H q$ and $q \bar{q} \to H g$ mediated by a nearly massless quark,''
  arXiv:1702.00426 [hep-ph].
  %%CITATION = ARXIV:1702.00426;%%

\bibitem{schmidt} C.~R.~Schmidt, 
Phys. Lett. {\bf B 413}, 391 (1997).

\bibitem{DelDuca}
%\cite{DelDuca:2001eu} 
  V.~Del Duca, W.~Kilgore, C.~Oleari, C.~Schmidt and D.~Zeppenfeld,
  %``Higgs + 2 jets via gluon fusion,''
  Phys.\ Rev.\ Lett.\  {\bf 87}, 122001 (2001);
  %doi:10.1103/PhysRevLett.87.122001
  %[hep-ph/0105129].
   V.~Del Duca, W.~Kilgore, C.~Oleari, C.~Schmidt and D.~Zeppenfeld,
  %``Gluon fusion contributions to H + 2 jet production,''
  Nucl.\ Phys.\ B {\bf 616}, 367 (2001).
  %doi:10.1016/S0550-3213(01)00446-1
  %[hep-ph/0108030].
  %%CITATION = doi:10.1103/PhysRevLett.87.122001;%%


%\cite{Neumann:2016dny}
\bibitem{Neumann:2016dny} 
  T.~Neumann and C.~Williams,
  %``The Higgs boson at high $p_T$,''
  Phys.\ Rev.\ D {\bf 95}, 014004 (2017).
%  doi:10.1103/PhysRevD.95.014004
%  [arXiv:1609.00367 [hep-ph]].
  %%CITATION = doi:10.1103/PhysRevD.95.014004;%%
  %5 citations counted in INSPIRE as of 06 Feb 2017




%\cite{Cascioli:2011va}
\bibitem{Cascioli:2011va} 
  F.~Cascioli, P.~Maierh\"{o}fer and S.~Pozzorini,
  %``Scattering Amplitudes with Open Loops,''
  Phys.\ Rev.\ Lett.\  {\bf 108}, 111601 (2012).
%  doi:10.1103/PhysRevLett.108.111601
%  [arXiv:1111.5206 [hep-ph]].
  %%CITATION = doi:10.1103/PhysRevLett.108.111601;%%
  %366 citations counted in INSPIRE as of 06 Feb 2017

\bibitem{openloops}
The {\sc OpenLoops} one-loop generator by F.~Cascioli, J.~Lindert,
  P.~Maierh{\"o}fer and S.~Pozzorini is publicly available at
  \url{http://openloops.hepforge.org}.

\bibitem{olrefs}
  S.~Kallweit, J.~M.~Lindert, P.~Maierh\"{o}fer, S.~Pozzorini and M.~Sch\"{o}nherr,
  %``NLO QCD+EW predictions for V + jets including off-shell vector-boson decays and multijet merging,''
  JHEP {\bf 1604}, 021 (2016);
  %doi:10.1007/JHEP04(2016)021
  %[arXiv:1511.08692 [hep-ph]].
  %%CITATION = doi:10.1007/JHEP04(2016)021;%%
  S.~H\"oche, P.~Maierh\"{o}fer, N.~Moretti, S.~Pozzorini and F.~Siegert,
  %``Next-to-leading order QCD predictions for top-quark pair production with up to three jets,''
  arXiv:1607.06934 [hep-ph].
  %%CITATION = ARXIV:1607.06934;%%

%\cite{Jezo:2016ujg}
\bibitem{Jezo:2016ujg} 
  T.~Jezo, J.~M.~Lindert, P.~Nason, C.~Oleari and S.~Pozzorini,
  %``An NLO+PS generator for $t \bar t$ and $W t$ production and decay including non-resonant and interference effects,''
  Eur.\ Phys.\ J.\ C {\bf 76}, no. 12, 691 (2016).
%  doi:10.1140/epjc/s10052-016-4538-2
%  [arXiv:1607.04538 [hep-ph]].
  %%CITATION = doi:10.1140/epjc/s10052-016-4538-2;%%
  %10 citations counted in INSPIRE as of 06 Feb 2017


%\cite{Cascioli:2014yka}
\bibitem{Cascioli:2014yka} 
  F.~Cascioli {\it et al.},
  %``ZZ production at hadron colliders in NNLO QCD,''
  Phys.\ Lett.\ B {\bf 735}, 311 (2014);
%  doi:10.1016/j.physletb.2014.06.056
%  [arXiv:1405.2219 [hep-ph]].
  %%CITATION = doi:10.1016/j.physletb.2014.06.056;%%
  %127 citations counted in INSPIRE as of 06 Feb 2017
%\cite{deFlorian:2016uhr}
%\bibitem{deFlorian:2016uhr} 
  D.~de Florian, M.~Grazzini, C.~Hanga, S.~Kallweit, J.~M.~Lindert, P.~Maierh\"{o}fer, J.~Mazzitelli and D.~Rathlev,
  %``Differential Higgs Boson Pair Production at Next-to-Next-to-Leading Order in QCD,''
  JHEP {\bf 1609}, 151 (2016);
 % doi:10.1007/JHEP09(2016)151
 % [arXiv:1606.09519 [hep-ph]].
  %%CITATION = doi:10.1007/JHEP09(2016)151;%%
  M.~Grazzini, S.~Kallweit, S.~Pozzorini, D.~Rathlev and M.~Wiesemann,
  %``$W^{+}W^{−}$ production at the LHC: fiducial cross sections and distributions in NNLO QCD,''
  JHEP {\bf 1608}, 140 (2016)
  %doi:10.1007/JHEP08(2016)140
  %[arXiv:1605.02716 [hep-ph]].
  %%CITATION = doi:10.1007/JHEP08(2016)140;%%

\bibitem{collier}
%\cite{Denner:2016kdg}
%\bibitem{Denner:2016kdg} 
  A.~Denner, S.~Dittmaier and L.~Hofer,
  %``Collier: a fortran-based Complex One-Loop LIbrary in Extended Regularizations,''
  Comput.\ Phys.\ Commun.\  {\bf 212}, 220 (2017);
 % doi:10.1016/j.cpc.2016.10.013
 % [arXiv:1604.06792 [hep-ph]].
  %%CITATION = doi:10.1016/j.cpc.2016.10.013;%%
  A.~Denner and S.~Dittmaier,
  %``Reduction of one loop tensor five point integrals,''
  Nucl.\ Phys.\ B {\bf 658}, 175 (2003);
 %% [hep-ph/0212259].
  %%CITATION = doi:10.1016/S0550-3213(03)00184-6;%%
  A.~Denner and S.~Dittmaier,
  %``Reduction schemes for one-loop tensor integrals,''
  Nucl.\ Phys.\ B {\bf 734}, 62 (2006);
 % doi:10.1016/j.nuclphysb.2005.11.007
 % [hep-ph/0509141].
  %%CITATION = doi:10.1016/j.nuclphysb.2005.11.007;%%
  A.~Denner and S.~Dittmaier,
  %``Scalar one-loop 4-point integrals,''
  Nucl.\ Phys.\ B {\bf 844}, 199 (2011).
  %doi:10.1016/j.nuclphysb.2010010.11.002
 % [arXiv:1005.2076 [hep-ph]].
  %%CITATION = doi:10.1016/j.nuclphysb.2010.11.002;%%

\bibitem{powheg}
%\cite{Nason:2004rx}
%\bibitem{Nason:2004rx} 
  P.~Nason,
  %``A New method for combining NLO QCD with shower Monte Carlo algorithms,''
  JHEP {\bf 0411}, 040 (2004);
 % doi:10.1088/1126-6708/2004/11/040
 % [hep-ph/0409146].
  %%CITATION = doi:10.1088/1126-6708/2004/11/040;%%
  S.~Frixione, P.~Nason and C.~Oleari,
  %``Matching NLO QCD computations with Parton Shower simulations: the POWHEG method,''
  JHEP {\bf 0711}, 070 (2007);
 % doi:10.1088/1126-6708/2007/11/070
 % [arXiv:0709.2092 [hep-ph]].
  %%CITATION = doi:10.1088/1126-6708/2007/11/070;%%
  S.~Alioli, P.~Nason, C.~Oleari and E.~Re,
  %``A general framework for implementing NLO calculations in shower Monte Carlo programs: the POWHEG BOX,''
  JHEP {\bf 1006}, 043 (2010);
  %doi:10.1007/JHEP06(2010)043
  %[arXiv:1002.2581 [hep-ph]].
  T.~Jezo and P.~Nason,
  JHEP {\bf 1512}, 065 (2015).


\bibitem{Frixione:1995ms} 
  S.~Frixione, Z.~Kunszt and A.~Signer,
  %``Three jet cross-sections to next-to-leading order,''
  Nucl.\ Phys.\ B {\bf 467}, 399 (1996).
  %doi:10.1016/0550-3213(96)00110-1
  %[hep-ph/9512328].
  %%CITATION = doi:10.1016/0550-3213(96)00110-1;%%

\bibitem{Harlander:2012hf} 
  R.~V.~Harlander, T.~Neumann, K.~J.~Ozeren and M.~Wiesemann,
  %``Top-mass effects in differential Higgs production through gluon fusion at order \alpha_s^4,''
  JHEP {\bf 1208}, 139 (2012).
%  doi:10.1007/JHEP08(2012)139
%  [arXiv:1206.0157 [hep-ph]].
  %%CITATION = doi:10.1007/JHEP08(2012)139;%%
 

%\cite{Ball:2014uwa}
\bibitem{Ball:2014uwa} 
  R.~D.~Ball {\it et al.} [NNPDF Collaboration],
  %``Parton distributions for the LHC Run II,''
  JHEP {\bf 1504}, 040 (2015).
%  doi:10.1007/JHEP04(2015)040
%  [arXiv:1410.8849 [hep-ph]].
  %%CITATION = doi:10.1007/JHEP04(2015)040;%%

\bibitem{rundec}
  K.~G.~Chetyrkin, J.~H.~Kuhn and M.~Steinhauser,
  %``RunDec: A Mathematica package for running and decoupling of the strong coupling and quark masses,''
  Comput.\ Phys.\ Commun.\  {\bf 133}, 43 (2000).
  %doi:10.1016/S0010-4655(00)00155-7
%  [hep-ph/0004189].
  %%CITATION = doi:10.1016/S0010-4655(00)00155-7;%%


\end{thebibliography}
\end{document}